\documentclass[twocolumn,showpacs,preprintnumbers,amsmath,amssymb]{revtex4}
\usepackage{graphicx}% Include figure files
\usepackage{dcolumn}% Align table columns on decimal point
\usepackage{bm}% bold math

\begin{document}

%\preprint{APS/123-QED}

\title{
Can one identify the intrinsic structure of the yrast states in $^{48}$Cr after the backbending?}
%Does the intrinsic state exist after backbending in $^{48}$Cr?}

\author{Zao-Chun Gao$^{1}$}%
\author{Mihai Horoi$^2$}
\author{Y. S. Chen$^1$}
\author{Y. J. Chen$^1$}
\author{Tuya$^3$}
\affiliation{
$^1$China Institute of Atomic Energy P.O. Box 275-10, Beijing 102413, China\\
$^2$Department of Physics, Central Michigan University, Mount Pleasant, Michigan 48859, USA\\
$^3$College of Physics Science and Technology, Shenyang Normal University, ShenYang, 110034, China }

\date{\today}% It is always \today, today,

\begin{abstract}
The backbending phenomenon in $^{48}$Cr has been investigated using the recently developed Projected Configuration Interaction (PCI) method, in which the
deformed intrinsic states are directly associated with shell model (SM) wavefunctions.
% KB3 interaction was used.
Two previous explanations, (i) $K=0$ band crossing,
%by the GCM (or CHFB in some sense),
and (ii) $K=2$ band crossing
% by PSM,
have been reinvestigated using PCI, and it was found that both explanations can successfully reproduce the experimental backbending. The PCI
wavefunctions in the pictures of $K=0$ band crossing and $K=2$ band crossing are highly overlapped. We conclude that there are no unique intrinsic states
associated with the yrast states after backbending in $^{48}$Cr.
\end{abstract}

\maketitle

\section{Introduction}

The backbending of $^{48}$Cr has been observed more than 10 years ago \cite{Cameron94,Cameron96}, but its interpretation
remains controversial and
challenges the existing nuclear models. Shell Model (SM) calculations have reproduced very well the yrast states of $^{48}$Cr \cite{Caurier94,Caurier95},
but it is difficult to provide the physical insight because the laboratory frame wavefunction doesn't imply any information associated with the deformed
intrinsic structure. The cranked Hartree Fock Bogoliubov(CHFB) method is a complementary theory \cite{chfb} often used to
analyze the deformed intrinsic states.
According to one CHFB analysis\cite{Caurier95}, $^{48}$Cr is an axial rotor up to the backbending, after that the system
changes to a spherical shape.
An alternative and more detailed
CHFB analysis \cite{Tanaka98} shows that the backbending of the $^{48}$Cr is not associated with the single particle level crossing and that the intrinsic configuration remains unchanged.

The Projected Shell Model (PSM) \cite{Hara95,Chen01} is an alternative technique that mixes the best intrinsic shell model configuration with other
associated particle-hole configurations. A PSM analysis \cite{Hara99} indicates that the backbending in $^{48}$Cr is due to  a band crossing involving an
excited 4-quasiparticle (qp) band with $K=2$, which represents a configuration of broken neutron and proton pairs. Unfortunately, such an explanation is
not confirmed by recent SM \cite{Caurier95,Bran05} and CHFB \cite{Caurier95,Tanaka98} analyses. In addition, the same PSM Ref. \cite{Hara99}, uses the
Generator Coordinate Method (GCM) that provides the picture of a spherical band crossing, which in some sense is similar to the result of CHFB.
Obviously, the above apparently conflicting explanations coming from different models need to be reinvestigated.

In the present paper, the backbending in $^{48}$Cr is investigated in the framework of the newly developed method called Projected Configuration Interaction (PCI)\cite{Gao09a,Gao09b}.
The PCI basis is built from a set of Slater determinants (SD). Those SDs may have different shapes, including the spherical shape.
Hence, the nuclear states with different intrinsic shapes can be mixed by the residual interaction.
%simultaneously obtained on the same footing.
By using the same SM Hamiltonian, the PCI results can be directly compared with those of full Shell Model calculations. Moreover, PCI uses deformed
single particle bases and therefore, the physics insight of the results can be clearly analyzed. Different PCI bases can be built in such a way that they
reflect the nature of the intrinsics states found in previous studies, such as CHFB (or GCM) and PSM. PCI wavefunctions were shown to be very good
approximations to those of full SM, and they can be obtained using different bases. Thereafter, overlaps among PCI wavefunctions can be calculated and
analyzed to determine the validity of various  explanations. These features suggest that PCI could shed new light on the interesting phenomenon of
backbending in $^{48}$Cr. Other models using similar techniques includes  the family of VAMPIR \cite{vampir}, and the quantum Monte Carlo diagonalization
(QMCD) method\cite{qmcd}.

The paper is organized as follows: Section II provides a short introduction into the basics concepts used by the PCI
method that would be required for an understanding of the arguments used later in the analysis. Section III is devoted
to the analysis of the contribution of different intrinsics set of states to the backbending in $^{48}$Cr.
Section IV summarizes the conclusion of our study.

\section{Basic concepts of the PCI method}

For completeness we give here a brief introduction of the PCI method (see Refs. \cite{Gao09a,Gao09b} for more details).
The deformed single particle (s.p.) states need to be generated either by HF calculation or from a deformed
s.p. Hamiltonian.
 For simplicity, here we take the latter approach, and the s.p. Hamiltonian can be written as
\begin{eqnarray}\label{hsp}
H_\text{s.p.}=h_\text{sph}
-\frac23\epsilon_2\hbar\omega_0 \rho^2P_2+\epsilon_4\hbar\omega_0 \rho^2P_4 ,
\label{sph}
\end{eqnarray}
where $h_\text {sph}=\sum_i e_i c^\dagger_i c_i$
 is the spherical s.p. Hamiltonian assumed to have
the same eigenfunctions as the spherical harmonic oscillator, $e_i$ energies are properly adjusted such that the SD with
lowest energy is close to the HF vacuum. For the $pf$-shell we use  $e_{f_{7/2}}=0.0$MeV, $e_{p_{3/2}}=4.5$MeV, $e_{f_{5/2}}=5.0$MeV
and $e_{p_{1/2}}=6.0$MeV.
 In Eq. (\ref{sph}) $\epsilon_2$, $\epsilon_4$ are the quadrupole and hexadecapole deformation parameters,
$P_l$ are Legendre polynomials, $\rho=r/b$, and we take $b=1.01A^{1/6}$ for the harmonic oscillator parameter \cite{Caurier94,Caurier95}.

The Slater determinants can be built with deformed s.p. states. Following our previous papers \cite{Gao09a,Gao09b},
the general structure of the PCI basis can be written as
\begin{eqnarray}\label{basis}
\left\{\begin{matrix}
  0{\text p}-0{\text h},  & \>  n{\text p}-n{\text h} \\
|\kappa_1,0\rangle, & |\kappa_1,j\rangle,\cdots, \\
|\kappa_2,0\rangle, & |\kappa_2,j\rangle,\cdots, \\
\hdotsfor{2}\\
|\kappa_N,0\rangle, & |\kappa_N,j\rangle,\cdots
\end{matrix}\right\},
\end{eqnarray}
where $|\kappa_i,0\rangle$ ($i=1,...N$) is an optimal \cite{Gao09b} set of starting states having different deformations.
Assuming that these $|\kappa,0\rangle$ are found (in what follows we skip the subscript $i$ to keep notation short),
a number of relative $n$p-$n$h SDs, $|\kappa,j\rangle$, on top of each
 $|\kappa,0\rangle$ are added to the SD basis selected the constraint \cite{Gao09a}
 \begin{eqnarray}\label{Ecut}
\Delta E=\frac12(E_0-E_j+\sqrt{(E_0-E_j)^2+4|V|^2})\geq E_\text{cut}.
\end{eqnarray}
Here $E_0=\langle\kappa,0|H|\kappa,0\rangle$, $E_j=\langle\kappa,j|H|\kappa,j\rangle$ and $V=\langle\kappa,0|H|\kappa,j\rangle$.
The PCI basis is then obtained by projecting the selected SDs onto good angular momentum. The wavefunctions, as well
as the energy levels, are obtained by solving the following generalized eigenvalue equation:
\begin{eqnarray}\label{eigen}
\sum_{\kappa'}(H_{\kappa \kappa'}^I-E^IN_{\kappa
 \kappa'}^I)f^I_{\kappa'}=0.
\end{eqnarray}
Here $H_{\kappa\kappa'}^I$ and $N_{\kappa \kappa'}^I$
are given by
\begin{eqnarray}\label{HN}
H_{\kappa\kappa'}^I=\langle
\kappa|HP^I_{KK'}|\kappa'\rangle, ~~ N_{\kappa
\kappa'}^I&=&\langle \kappa|P^I_{KK'}|\kappa'\rangle,
\end{eqnarray}
where $P^I_{KK'}$ is the angular momentum projection operator, and
$H$ is the shell model Hamiltonian. In this study we take the KB3 interaction \cite{Poves81}, which has
been used by Caurier et al in their shell model calculations of $^{48}$Cr \cite{Caurier95}. However,
results similar to those reported here are provided by other interactions, such as KB3G \cite{kb3g} or GXPF1A \cite{gxpf1a}.

\section{The PCI analysis of the backbending in $^{48}$Cr}

\begin{figure}
\centering
\includegraphics[width=3.5in]{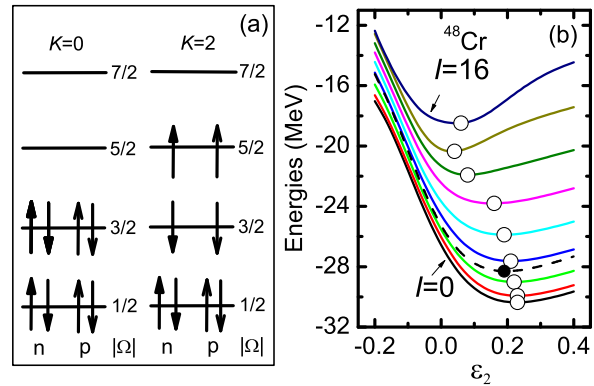}
\caption{ (color online) (a) The $K=0$ and $K=2$ configurations used in the present calculations of $^{48}$Cr. All 4 levels are from $1f_{7/2}$ subshell.
(b) Potential energy curves of the $K=0$ configuration in (a) as functions of $\epsilon_2$. The solid lines show the projected energies and the dashed
line, unprojected energy. $\epsilon_4$ was chosen to minimize the potential energy for each $\epsilon_2$. KB3 interaction was used. } \label{sd}
\end{figure}

To get some insight into the structure of the states contributing to the backbending in $^{48}$Cr
we chose two basic $|\kappa,0\rangle$ SDs. The first $|\kappa_1,0\rangle$ is a $K=0$ configuration with all 8 valence
nucleons occuping the $|\Omega|=1/2(1f_{7/2})$ orbits and the $|\Omega|=3/2(1f_{7/2})$ orbits, as shown in Fig. \ref{sd}a. The deformation of
$|\kappa_1,0\rangle$ is given by $\epsilon_2=0.19$ and $\epsilon_4=-0.05$.
This deformation was obtained by determining the minimum of the energy surface of $\langle
K=0|H_\text{KB3}|K=0\rangle$ as a function of $\epsilon_2$ and $\epsilon_4$, (See the dashed line in Fig. \ref{sd}b). A minimum energy of
$\langle\kappa_1,0|H_\text{KB3}|\kappa_1,0\rangle=-28.296$ MeV was found, which is close to the HF energy $-28.423$ MeV.
$|\kappa_1,0\rangle$ is believed to be responsible for the low-spin yrast states before the backbending, and the corresponding PCI basis
that includes the $n$p-$n$h states selected by Eq. (\ref{Ecut}) is
denoted as `g.s.', and it is shown in Table \ref{SDs}.

\begin{table}
\caption{\label{SDs} PCI bases used for the backbending study in $^{48}$Cr. $K=0$ and $K=2$ configurations are shown in Fig. \ref{sd}.}
\begin{ruledtabular}
\begin{tabular}{c|ccc|ccc|ccc|ccc}
Basis&&g.s.&&&A&&& B&& & C&\\
\hline
     &  $K$&$\epsilon_2$&$\epsilon_4$ &  $K$&$\epsilon_2$&$\epsilon_4$ &$K$&$\epsilon_2$&$\epsilon_4$ &$K$&$\epsilon_2$&$\epsilon_4$\\
$|\kappa,0\rangle$& 0&0.19& $-0.05$& 0&0.00&  0.00&2&0.00& 0.00&2&0.19&$-0.05$\\
$N_\kappa$& &143&  &&192& &&186& &&149&\\
\end{tabular}
\end{ruledtabular}
\end{table}

The second basic SD, $|\kappa_2,0\rangle$, is chosen to describe the high spin states in $^{48}$Cr, after the backbending.
According to the previous studies mentioned in the introduction, there are at least two candidate configurations for $|\kappa_2,0\rangle$.
The first, suggested by the CHFB calculations, according to which the backbending in $^{48}$Cr can be explained
without a band crossing \cite{Tanaka98}, but by a shape changes from well deformation to spherical \cite{Caurier95}.
Our PCI calculations seem to be in agreement with this interpretation. The projected energy curves for the $K=0$ configuration at each even-spin
are shown in Fig. \ref{sd}b (solid lines).
One can see that the deformation at minimum decreases as the spin increases.
Guided by this result, one can establish a possible $|\kappa_2,0\rangle$, which has the same configuration as $|\kappa_1,0\rangle$,
but whose shape is almost spherical.
Such choice of $|\kappa_2,0\rangle$ SD would be consistent with the GCM interpretation, in which HF vacua with different deformations are included in the basis.
However, in PCI one can further include particle-hole excitations on top of each $|\kappa,0\rangle$.
On should recall that in the PSM interpretation \cite{Hara99} the
backbending in $^{48}$Cr is caused by a $K=2$ band crossing.
Therefore, PCI seem to be ideally suited to include and analyze another possible $|\kappa_2,0\rangle$ with $K=2$.
Its configuration is also shown in Fig. \ref{sd}a, and its structure can be chosen either spherical
(see column B in Table \ref{SDs}) or of the same deformation as that of
$|\kappa_1,0\rangle$ favored by the PSM approach (see column C in Table \ref{SDs}).
In an attempt to find the optimal structure of yrast states in $^{48}$Cr after the backbending,
we considered all three possibilities of $|\kappa_2,0\rangle$, labeled with `A',
`B' and `C' in Table \ref{SDs}.

The $n$p-$n$h $|\kappa,j\rangle$ SDs built on top of each $|\kappa,0\rangle$ are selected by
setting $E_\text{cut}=0.5$ keV in Eq. (\ref{Ecut}).
Consequently, the number of selected $|\kappa_1,j\rangle$ is 142, and those of $|\kappa_2,j\rangle$ for A, B and C
are 191, 185 and 148, respectively.
Adding the $|\kappa,0\rangle$ itself, the total number of the selected SDs, $N_\kappa$, for each $|\kappa,0\rangle$ is
  listed in Table \ref{SDs}.

\begin{figure}
\centering
\includegraphics[width=2.5in]{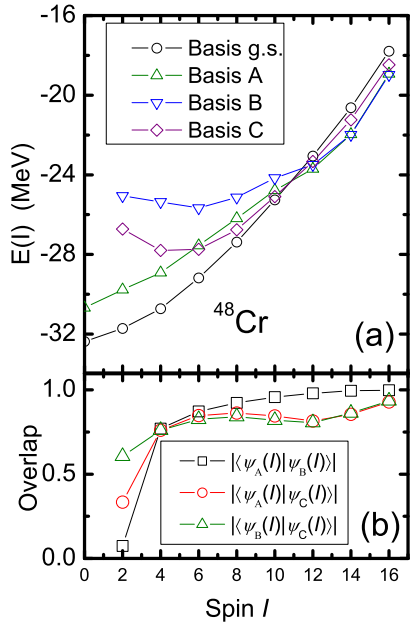}
\caption{(color online) (a) PCI energies $E(I)$ as functions of spin for the bases listed in Table \ref{SDs}.
(b) Overlaps among the PCI wavefunctions for bases A, B, and C.} \label{Cr48a}
\end{figure}

\begin{figure}
\centering
\includegraphics[width=2.5in]{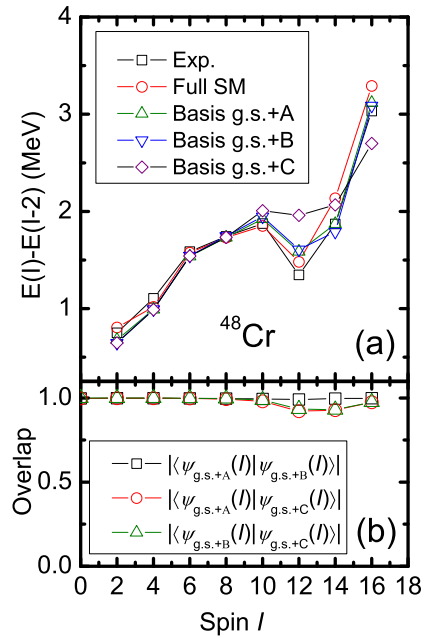}
\caption{(color online) (a) $E2$ transition energies $E(I)-E(I-2)$ vs spin $I$. Experimental data are taken from Ref. \cite{Cameron96}, and the result of
the Full SM from Ref. \cite{Caurier95}. (b) Overlaps among the PCI wavefunctions with basis sets g.s.+A, g.s.+B and g.s.+C. } \label{Cr48b}
\end{figure}

These PCI bases were used to calculate all even-spin states from $I=0$ to $I=16$. The calculated lowest energies with basis g.s., A, B and C as functions
of spin are shown in Fig. \ref{Cr48a}(a). The energies calculated with basis g.s. alone increases very smoothly with the spin, without exhibiting any
backbending. For the bases A, B and C, all the corresponding energy curves cross that of the g.s. band after spin $I=10$. To get a better understanding
of the relation among those bases, we calculated the overlaps between corresponding wavefunctions. Let's denote the PCI wavefunctions as
$|\Psi_\kappa(I)\rangle$, where $\kappa$ refers to certain basis or combination. For instance, $\kappa$ can be  `g.s.', `A', or `g.s.+A', etc. Overlaps
of $|\langle\Psi_A(I)|\Psi_B(I)\rangle|$, $|\langle\Psi_A(I)|\Psi_C(I)\rangle|$ and $|\langle\Psi_B(I)|\Psi_C(I)\rangle|$ as functions of $I$ are plotted
in Fig. \ref{Cr48a}(b). The surprising result is that these overlaps are unexpectedly very large, which means quite different intrinsic bases would
generate almost the same wavefunctions after the angular momentum projection. In particular, one should note the large overlap
$|\langle\Psi_A(I)|\Psi_B(I)\rangle|\approx1$ at $I>10$, while without the angular momentum projection bases A and B are strictly orthogonal due to
different $K$ values .

The $E_\gamma(I)=E(I)-E(I-2)$ energies calculated using combinations of bases, g.s.+A, g.s.+B and g.s.+C
are shown in Fig. \ref{Cr48b}(a).
The backbending phenomenon seems to be easily reproduced by all those bases.
As already discussed, the g.s.+A basis is qualitatively similar to that used in the CHFB and GCM analyses, while the g.s.+C basis follows the scenario proposed by the PSM analysis.
%Therefore, thus our result also supports the PSM.
Furthermore, as shown in Fig. \ref{Cr48b}(b), the overlaps $|\langle\Psi_{g.s.+A(I)}|\Psi_{g.s.+C}(I)\rangle|$ are very large,
at least $92\%$.
Therefore, we come to the conclusion that the apparently contradictory CHFB and
PSM explanations, actually look qualitatively equivalent one to each other.
% However, quantitatively,
One should note, however, that the backbending described with the g.s.+C basis is quantitatively not as good as that descibed
with the g.s.+A basis. The reason could be the large deformation of C basis.
Changing the deformation of C basis to spherical, one gets the g.s.+B basis, and the result is improved.
This feature supports the idea that the shape of $^{48}$Cr reduces
after backbending, as has been pointed out in Ref. \cite{Caurier95, Tanaka98}.

Notice that the results obtained with bases g.s.+A and g.s.+B are almost identical, although the intrinsic bases A and B have a completely different
structure. Our calculations show that there are also other intrinsic bases with $K$ different of 0 and 2 that can reproduce the backbending, and whose
corresponding wavefunctions are almost equivalent to the basis g.s.+A or g.s.+B when projected on good angular momenta. In other words, one can not find
a unique intrinsic state for the yrast states in $^{48}$Cr for $I=12-16$. One can get some insight into the apparent irrelevance of the intrinsic
structure at high spins by analyzing the case of $I=16$, which is the band termination state. In the space of $\pi 1f^4_{7/2} \nu 1f^4_{7/2}$, there is
only one SD that reaches the maximum $K=16$, showing that only one $I=16$ state can be constructed from that space. On the other hand many $\pi
1f^4_{7/2} \nu 1f^4_{7/2}$ SDs with various $K$ values
 can be projected onto good $I$. Therefore, we have many projected states with $I=16$,
 and our calculations prove that they are exactly identical and the projected energy is $-18.342$ MeV with KB3 interaction.
Therefore, one can use any one of the $\pi 1f^4_{7/2} \nu 1f^4_{7/2}$ SDs to reproduce the state of band termination.

\begin{figure}
\centering
\includegraphics[width=2.5in]{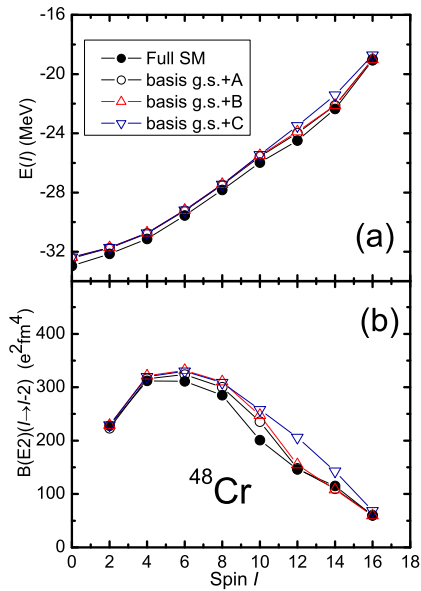}
\caption{(color online) (a) Yrast state energies vs spin obtained with PCI (open symbols) and Full SM(solid symbols).
(b) The BE2 values with the same wavefunctions as in (a). Result of the Full SM are taken from Ref. \cite{Caurier94}.\label{BE2}}
\end{figure}

\begin{figure}
\centering
\includegraphics[width=2.5in]{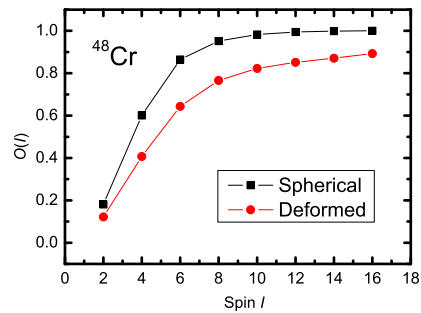}
\caption{ (color online) Overlaps O(I) in Eq.\ref{oi} between the projected intrinsic configurations with $K=0$ and $K=2$
 shown in Fig.\ref{sd}a for spherical and deformed ($\epsilon_2=0.19$ and $\epsilon_4=-0.05$) cases. } \label{ovlp}
\end{figure}

Fig. \ref{Cr48b} shows good comparison of the $E2$ transition energies obtained with the particular choice of bases described above and those of the full
SM calculations. One would like to further compare and validate the structure of different approaches with the full SM results. It is difficult to
calculate the overlap between PCI and full SM wavefunctions because the large number of norm matrix elements, $N_\text{PCI}\times N_\text{SM}$ that has
to be calculated (see Eq. (\ref{HN})).  Here $N_\text{PCI}$ and $N_\text{SM}$ are dimensions of the PCI and full (M-scheme) SM spaces, where
$N_\text{SM}$ is a very large number. However, the PCI approximation can be further checked by comparing the energies as well as the BE2 transition
probabilities with those of the full SM calculations. The results are shown in Fig. \ref{BE2}. The PCI energies lie 500 keV on average above the exact
values, but they were calculated using less than 400 SDs (see Table. \ref{SDs}), while in full SM calculations the number of the SDs used is 1963461. For
the  B(E2) calculations we used the `canonical' effective charges, i.e. $1.5e$ for protons and $0.5e$ for neutrons, which are the same as those used in
Refs. \cite{Caurier94,Caurier95}. The BE2 values obtained with PCI are very close to the exact ones, except those for $I=6,8$ and 10, which are slightly
larger. That is very likely due to the fixed deformation we used for the selected `g.s' basis. As shown in Fig. \ref{sd}b, the deformation reduces
gradually as the spin increases, conclusion also supported by the CHFB \cite{Caurier95} and the cranked Nilsson-Strutinsky (CNS) \cite{Juodagalvis06}
calculations. Therefore we overestimated the BE2 values at $I=6,8$ and 10.

%It is known that the occurrence of backbending in the rare-earth region is due to the broken pairs occupying the intruder high-$j$ orbits. The extra
%angular momentum of those broken pairs aligned with the collective rotation, leads to a sudden increase of the total angular momentum. Consequently, the
%backbending appears. However, in the case of $^{48}$Cr, there is no intruder orbits, the alignment can only be provided by the particles in the
%$1f_{7/2}$ orbits. Since $N=Z$, both neutron pair and proton pair may be broken simultaneously. Then the total alignment of those broken pairs should be
%$i=12$, which is actually the band termination in the $\pi 1f^2_{7/2} \nu 1f^2_{7/2}$ space\cite{Bran05,Juodagalvis98,Zuker95}. However, such $I=12$ can
%only be reached when the $f_{7/2}$ orbits is pure, that means the intrinsic shape becomes spherical.

On a final note the intrinsic states play a key role in studying the physics of the nuclear system. Even if they are orthogonal,
their overlap can be quite large after the angular momentum projection. Here is a simple example.
The $K=0$ and $K=2$ configurations in Fig. \ref{sd}a were projected on to good angular momentum assuming
 spherical and deformed ($\epsilon_2=0.19$ and $\epsilon_4=-0.05$) shapes, respectively. The overlap
\begin{eqnarray}\label{oi}
O(I)=\frac{\langle K=0|P^I_{02}|K=2\rangle}{\sqrt{\langle K=0|P^I_{00}|K=0\rangle\langle K=2|P^I_{22}|K=2\rangle}}
\end{eqnarray}
was calculated and plotted in Fig. \ref{ovlp}.
One can see that the deformed overlap $O(I)$ is always smaller than the spherical one, indicating that the deformed
intrinsic states can be more clearly defined than the spherical ones.
In the low-spin region the overlaps $O(I)$ are small, and the two intrinsic configurations can be easily distinguished.
But at high spin, $O(I)$ are much larger for both spherical and deformed cases, which means that the intrinsic states can not be clearly
identified, at least for the case of $^{48}$Cr.

\section{Summary}

In summary, the backbending in $^{48}$Cr has been studied with a recently developed Projected Configuration Interaction (PCI) method. PCI uses the same
realistic Hamiltonians and valence spaces as the SM calculations, but only a set of properly selected SDs with different deformations and associated
$n$p-$n$h configurations. The backbending in $^{48}$Cr has been reproduced by using various PCI bases, carefully selected to reflect the physics of the
intrinsic states found by the CHFB (GCM) and PSM analyses of this case. Using the PCI capabilities of mixing these bases we show for the first time that
the backbending pictures proposed by the CHFB and PSM methods are qualitatively equivalent. Our analysis supports the conclusion that there is no unique
intrinsic state for spins larger than 10 in $^{48}$Cr.

\vspace*{0.5cm}

This work is supported by the NSF of China Contract No. 10775182. Z.G. and M.H. acknowledge support from the US DOE UNEDF Grant No. DE-FC02-09ER41584.
M.H. acknowledges support from NSF Grant No. PHY-0758099. Y.S.C. acknowledges support from the MSBRDP of China under Contract No. 2007CB815003.

\end{document}